\documentclass[twocolumn]{aastex62}

\usepackage{soul}

\graphicspath{{./}{images/}}

\newcommand{\lat}{\textit{Fermi}/LAT}
\newcommand{\gbm}{\textit{Fermi}/GBM}
\newcommand{\xrt}{\textit{Swift}/XRT}
\newcommand{\bat}{\textit{Swift}/BAT}

\usepackage{comment}
\usepackage{amsmath}
\usepackage{multirow}
\usepackage{xcolor}

\begin{document}
\title{Constraints on the physics of the prompt emission from a distant and energetic $\rm \gamma$-ray burst GRB 220101A}
\correspondingauthor{Alessio Mei}
\email{alessio.mei@gssi.it}
\author{Alessio Mei}
\affiliation{Gran Sasso Science Institute, Viale F. Crispi 7,I-67100,L’Aquila (AQ), Italy}
\affiliation{INFN - Laboratori Nazionali del Gran Sasso, I-67100, L’Aquila (AQ), Italy}

\author{Gor Oganesyan}
\affiliation{Gran Sasso Science Institute, Viale F. Crispi 7,I-67100,L’Aquila (AQ), Italy}
\affiliation{INFN - Laboratori Nazionali del Gran Sasso, I-67100, L’Aquila (AQ), Italy}

\author{Anastasia Tsvetkova}
\affiliation{Ioffe Institute, Politekhnicheskaya 26, St. Petersburg 194021, Russia}

\author{Maria Edvige Ravasio}
\affiliation{Department of Astrophysics/IMAPP, Radboud University, P.O. Box 9010, 6500 GL, Nijmegen, The Netherlands}
\affiliation{INAF – Astronomical Observatory of Brera, via E. Bianchi 46, I-23807 Merate, Italy}

\author{Biswajit Banerjee}
\affiliation{Gran Sasso Science Institute, Viale F. Crispi 7,I-67100,L’Aquila (AQ), Italy}
\affiliation{INFN - Laboratori Nazionali del Gran Sasso, I-67100, L’Aquila (AQ), Italy}

\author{Francesco Brighenti}
\affiliation{Gran Sasso Science Institute, Viale F. Crispi 7,I-67100,L’Aquila (AQ), Italy}

\author{Samuele Ronchini}
\affiliation{Gran Sasso Science Institute, Viale F. Crispi 7,I-67100,L’Aquila (AQ), Italy}
\affiliation{INFN - Laboratori Nazionali del Gran Sasso, I-67100, L’Aquila (AQ), Italy}

\author{Marica Branchesi}
\affiliation{Gran Sasso Science Institute, Viale F. Crispi 7,I-67100,L’Aquila (AQ), Italy}
\affiliation{INFN - Laboratori Nazionali del Gran Sasso, I-67100, L’Aquila (AQ), Italy}

\author{Dmitry Frederiks}
\affiliation{Ioffe Institute, Politekhnicheskaya 26, St. Petersburg 194021, Russia}


\begin{abstract}
The emission region of $\rm \gamma$-ray bursts (GRBs) is poorly constrained. The uncertainty on the size of the dissipation site spans over 4 orders of magnitude ($\rm 10^{12}-10^{17}$ cm) depending on the unknown energy composition of the GRB jets. The joint multi-band analysis from soft X-rays to high energies (up to $\rm \sim$ 1 GeV) of one of the most energetic and distant GRB 220101A (z = 4.618) allows us for an accurate distinction between prompt and early afterglow emissions. The enormous amount of energy released by GRB 220101A ($\rm E_{iso} \approx 3 \times10^{54}$ erg) and the spectral cutoff at $\rm E_{cutoff} = 85_{-26}^{+16}$ MeV observed in the prompt emission spectrum constrains the parameter space of GRB dissipation site. We put stringent constraints on the prompt emission site, requiring $\rm 700<\Gamma_0<1160 $ and $\rm R_\gamma \sim 4.5 \times 10^{13}$ cm. Our findings further highlights the difficulty of finding a simple self
consistent picture in the electron-synchrotron scenario, favoring instead a proton-synchrotron model, which is also consistent with the observed spectral shape. Deeper measurements of the time variability of GRBs together with accurate high-energy observations (MeV-GeV) would unveil the nature of the prompt emission.
\end{abstract}

\section{Introduction} \label{sec:intro}
Despite many years of observations, the energy composition of $\rm \gamma$-ray burst (GRB) jets and the dissipation processes responsible for the production of the prompt emission remain open mysteries (e.g. see \citealt{Piran2004,Kumar2015} for a review). Models predicting the release of the prompt emission at the photosphere (e.g. \citealt{Rees2005,Peer2008}), via internal shocks \citep{Rees1994} or magnetic re-connection (e.g. \citealt{Spruit2002,Zhang2011}) are indistinguishable by using the current GRB observations.\\
Some GRB spectra have been successfully fitted by the slow-cooling \citep{Tavani1996} or by the self-absorbed \citep{Lloyd2000} synchrotron model. Two-component, non-thermal plus thermal component have been invoked to explain time-resolved spectra of few GRBs \citep{Burgess2014,Yu2015}, where an empirical function with fixed synchrotron spectral indices have been adopted. It was found, however, that GRB time-resolved and time-integrated spectra of GRBs can be well described by single non-thermal emission component with a low-energy spectral break \citep{Oganesyan2017,Oganesyan2018,Ravasio2019}, and corresponding spectral indices that are consistent with marginally fast cooling regime of the synchrotron radiation. It was also shown that the realistic, physically derived synchrotron radiation model can account for GRB spectra in a slow or marginally fast cooling regimes \citep{Oganesyan2019,Burgess2020}\footnote{More complete list of references, including empirical and physical modelling of single GRB spectra can be found in \cite{Zhang2020}} without the necessity to invoke additional thermal components.
However, some time-resolved spectra within a GRB are harder than predicted in the synchrotron radiation model (e.g. \citealt{Acuner2020}). While on one side it seems difficult to assign the synchrotron origin to all the GRB spectra, on the other hand it is quite clear that the presence of the high energy power law segment in the GRB spectra requires non-thermal radiative processes to be present. Moreover, the exact regime of the radiation does not directly correspond to the unique dissipation model. For instance, single-shot accelerated electrons in low magnetised ejecta \citep{Kumar2008,Beniamini2013}, re-accelerated electrons in highly magnetised ejecta \citep{Gill2020} or protons in the magnetically dominated jets \citep{Ghisellini2020} can produce the very same marginally fast cooling synchrotron spectra. Therefore, more specific observational inputs are required to discriminate between GRB jet dissipation models.\\
Regardless of the dominant radiative processes responsible for the GRB production, there should be a critical energy in the GRB spectrum above which the photons are suppressed by the pair production \citep{Ruderman1975,Piran1999}. The localisation of the high energy spectral cutoff $\rm E_{cutoff}$ enables to constrain the size of the jet $\rm R_{\gamma}$ as a function of the bulk Lorentz factor $\rm \Gamma_{0}$ where the prompt emission is produced \citep{Lithwick2001,Gupta2008,Granot2008,Zhang2009,Hascoet2012,Vianello2018,Chand2020}. Given the large typical $\rm \Gamma_0 \ge 100$ \citep{Ghirlanda2018}, the observed spectral cutoff results to be $\rm E_{cutoff} \ge (51\ MeV) \, \Gamma_{0,2} \, (1+z)^{-1}$. At these energies, the identification of the cutoff faces two complications: the extremely low instrumental response of operating GRB detectors, and the presence of an afterglow emission that typically over-shines the prompt emission at $0.1-10$ GeV for the majority of the GRBs detected by \lat\ (see \citealt{Nava2018} for a review).
\begin{figure*}
\centering 
	\includegraphics[width=0.45\textwidth]{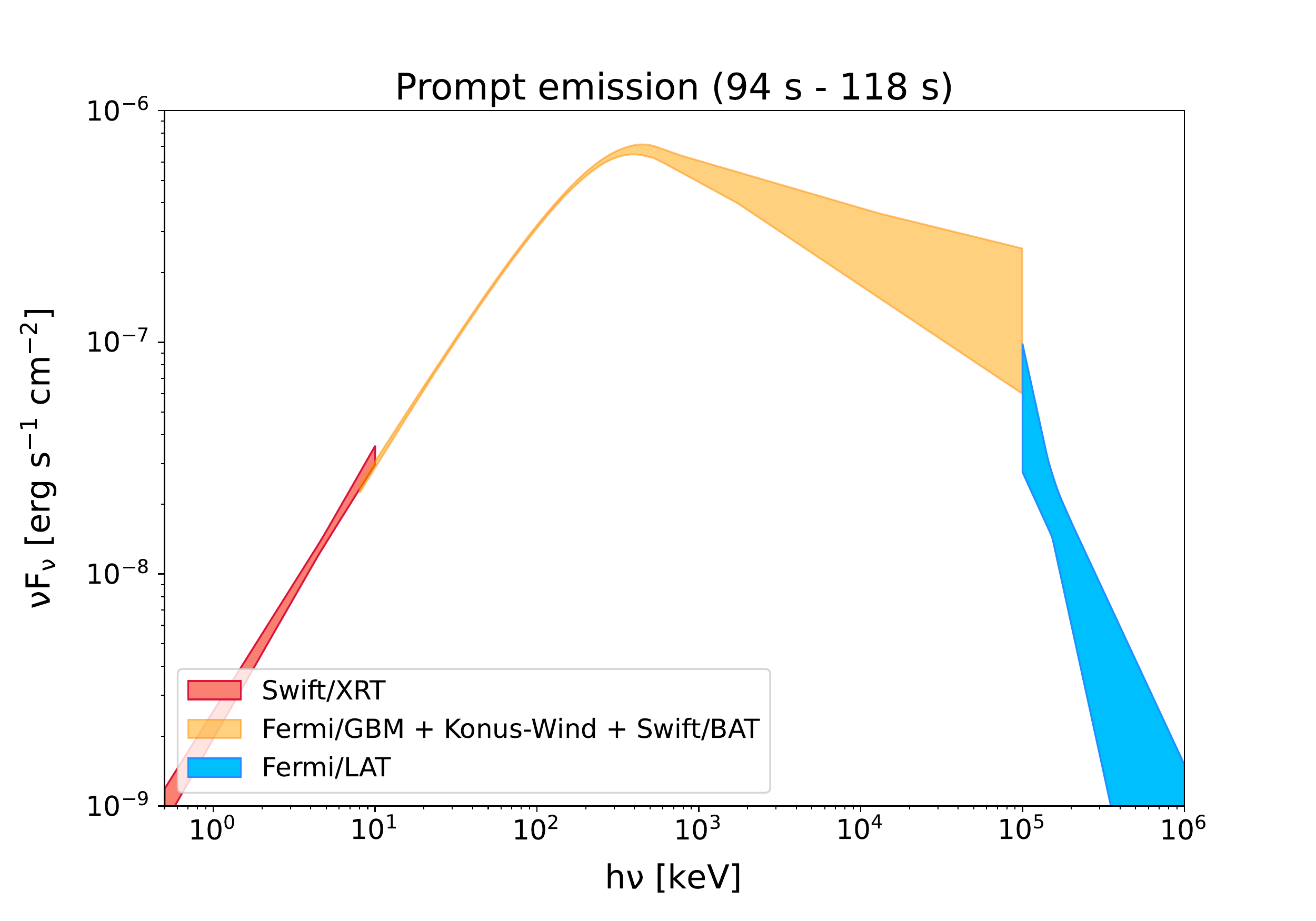}  \includegraphics[width=0.45\textwidth]{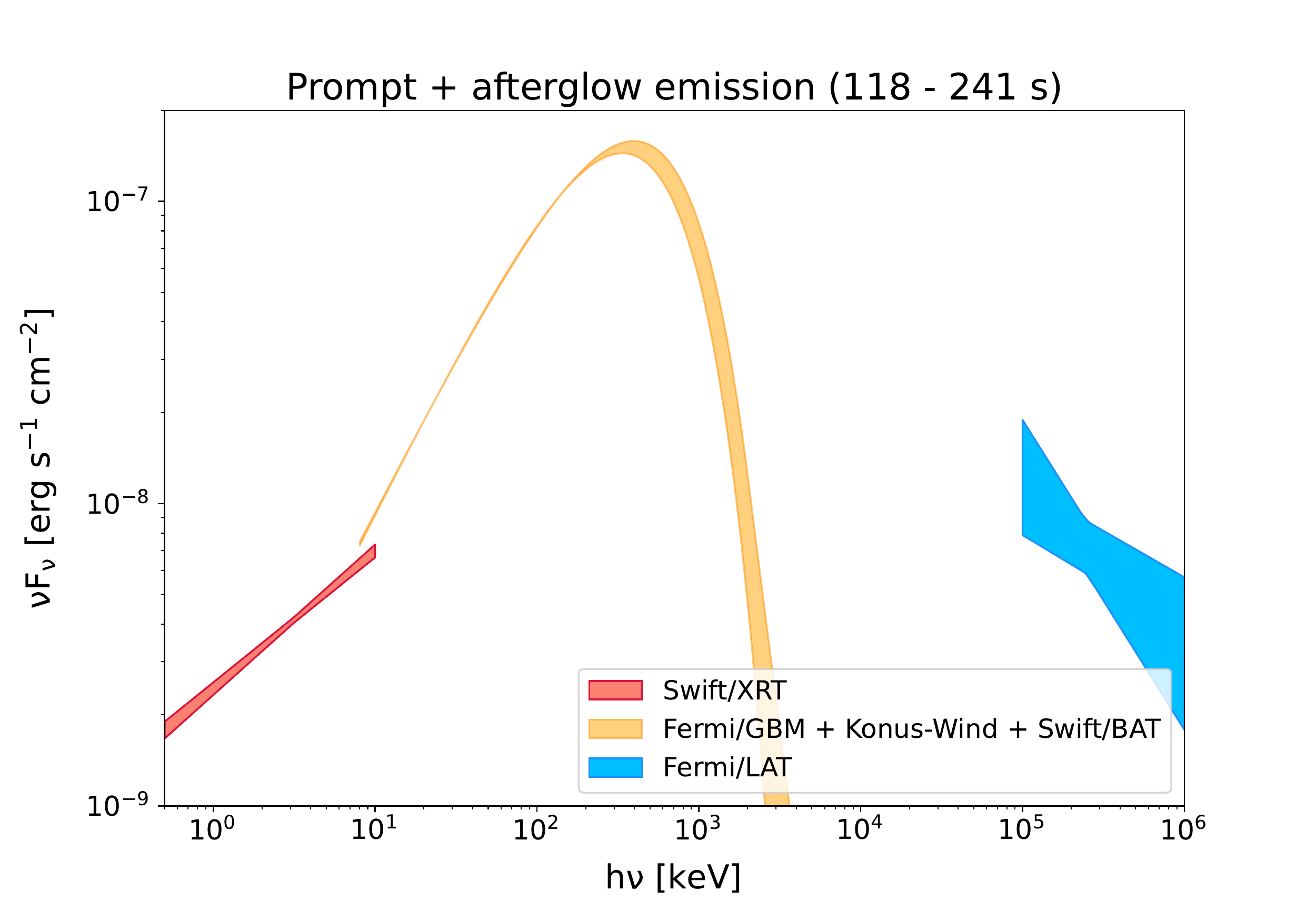}  
	\includegraphics[width=0.75\textwidth]{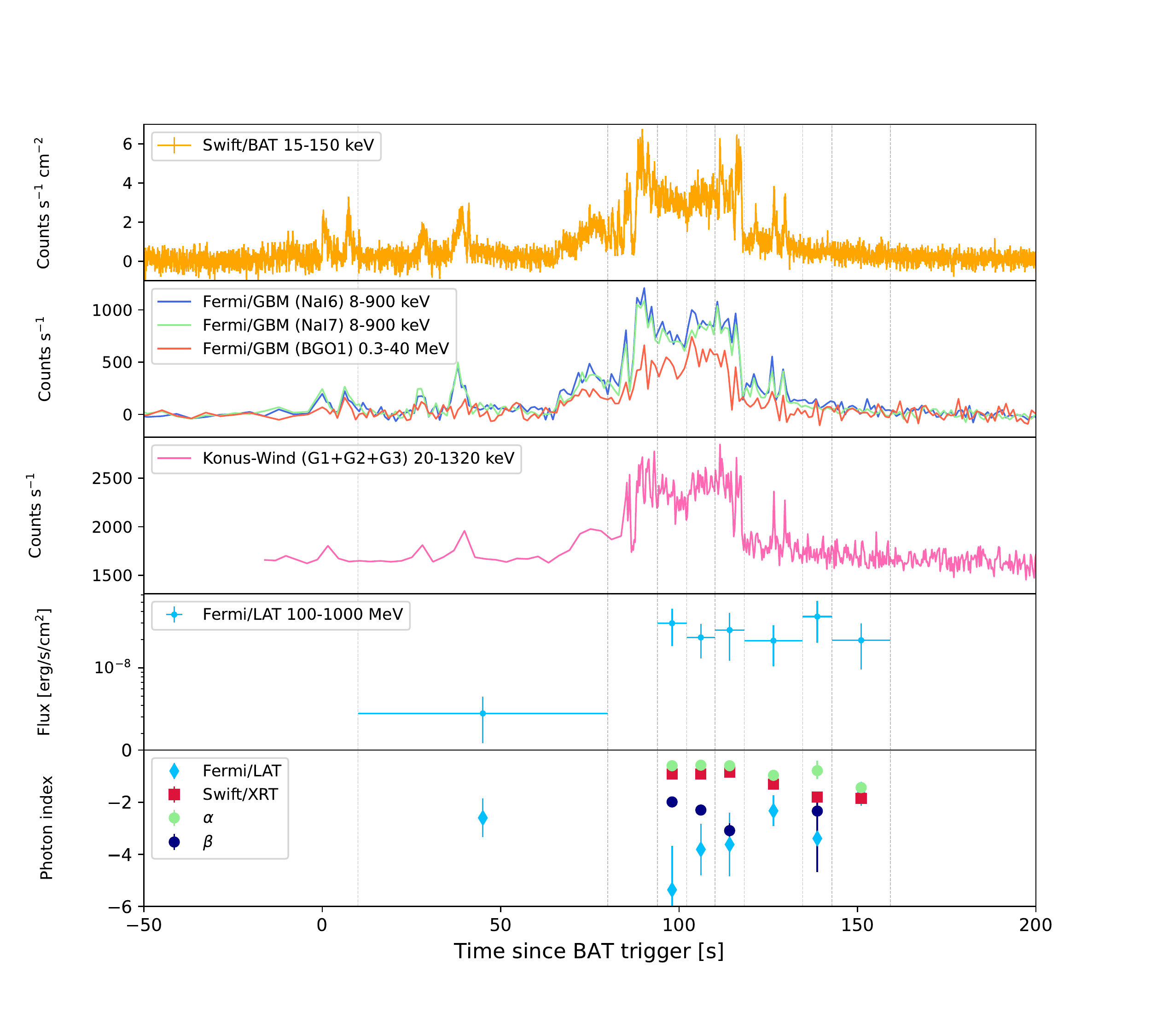}  
    \caption{\textit{Top Panel:} Modelled spectra considering three independent dataset: XRT (in red), BAT+GBM+KW (in orange) and LAT (in blue). The models are shown in the $\rm \nu F_{\nu}$ representation. Two distinguished time-epochs are chosen: the first one dominated fully by the prompt emission (94-118 s, on the left) and the second one (118-241 s, on the right) with the separate prompt emission (8 keV - 100 MeV) and afterglow components (in X-rays and high energies). We draw the energy-dependent flux uncertainty regions ("butterflies") derived from \textit{Swift}/XRT and \textit{Fermi}/LAT data analysis in the red and blue, respectively.\\
    \textit{Bottom Panel:} \textit{Swift}/BAT (15-150 keV), \gbm\ (8 keV-40 MeV), \textit{Konus}-WIND (13-750 keV) count rate light-curves, \textit{Fermi}/LAT flux light-curve  (0.1-1 GeV) and photon indices measured from the \textit{Swift}/XRT (0.5-10 keV) and \textit{Fermi}/LAT (0.1-1 GeV) time-resolved spectra.}
	\label{fig:spectra+LCs}
\end{figure*}
We analyse the broad-band data from soft X-rays (0.5-10 keV) to high energies ($\rm \sim 1$ GeV) of GRB 220101A, one of the most energetic GRB ($\rm E_{iso} \approx 3 \times 10^{54}$ erg) located at very high redshift (z = 4.618) and detected by the X-Ray Telescope (XRT, 0.5 - 10 keV) and Burst Alert Telescope (BAT, 15-150 keV) on-board the Neil Gehrels Swift Observatory ({\it Swift}), the Gamma-ray Burst Monitor (GBM, 8 keV - 40 MeV) and Large Area Telescope (LAT, 100 MeV - 300 GeV) on-board the {\it Fermi} satellite and Konus-{\it Wind} (KW, $\sim$20 keV - 20 MeV) instrument, together with several optical instruments. We identify the high energy spectral cutoff and place the most stringent constraints on the $\rm R_{\gamma}-\Gamma_{0}$ plane. These constraints are fully consistent with the limits obtained from the optical-to-GeV afterglow emission modelled by the dissipation of the jet in the circum-burst medium \citep{Paczynski1993}. We discuss the physical implications of our findings and stress on the necessity of better early MeV-GeV observations.\\

\section{Independent spectral analysis}\label{sec:indep_analysis}
\begin{figure*}
\centering 
    \includegraphics[width=0.7\textwidth]{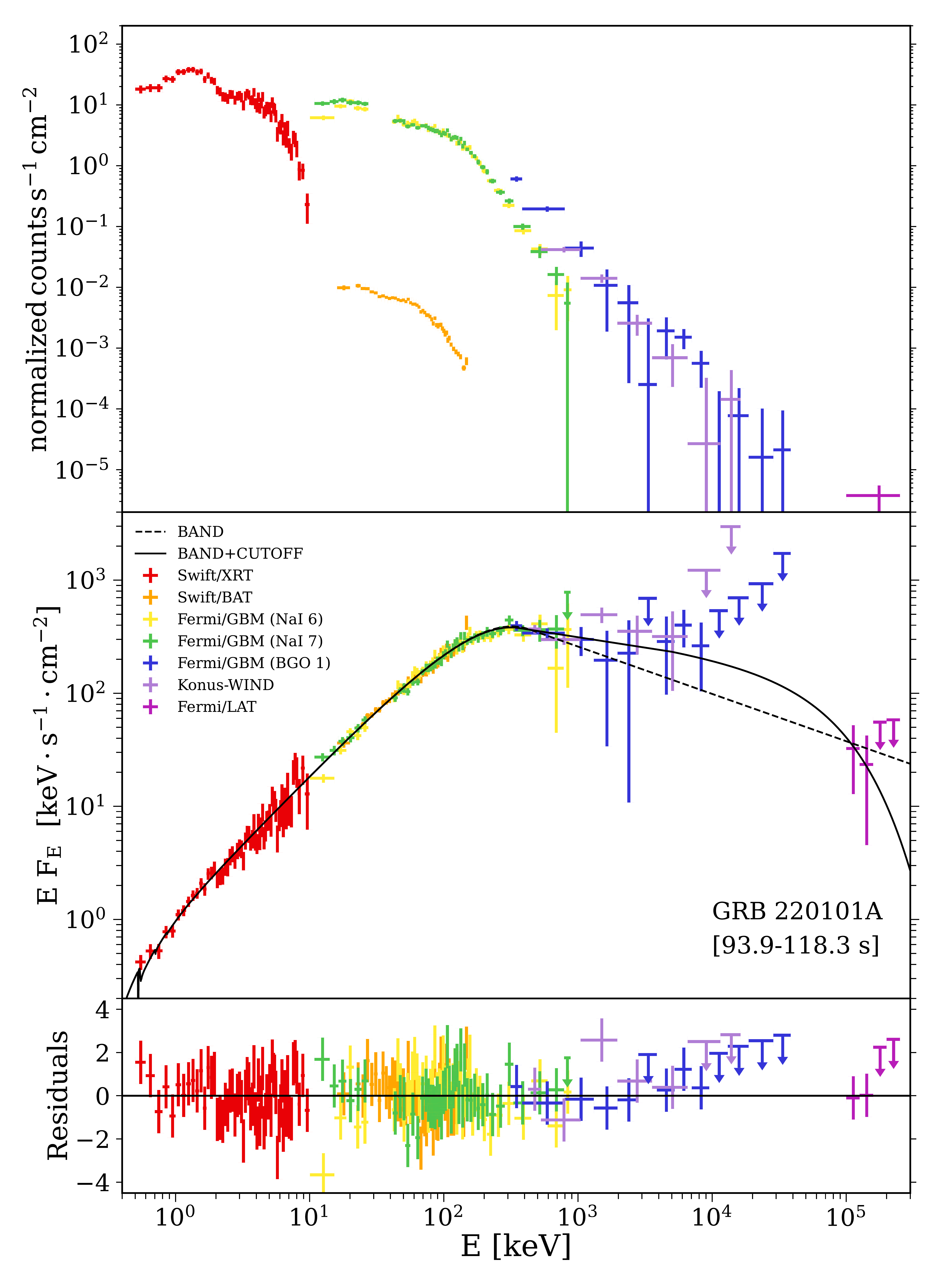}
    \caption{The joint X-ray to high-energy spectrum of GRB 220101A at 94 - 118 s modelled by Band function with the high energy spectral cutoff $\rm E_{cutoff}$ (\textit{Band+cutoff}). 
    We show the spectrum in the count representation (\textit{upper panel}), in the spectral energy density representation (SED, \textit{middle panel}) and the residuals between the data and best-fit model (\textit{bottom panel}). We show the data points and relative errors with different colors, depending on the instruments, while the 3$\sigma$ upper limits are shown with arrow markers. Joint \textit{Band} model fit is shown with a black-dashed line.}
	\label{fig:cutoff_prompt}
\end{figure*}
\begin{table*}
\begin{center}
\begin{tabular}{c|c|c|c|c|c|c}
\hline\hline
 Time bin [s]& Instruments  & $ \rm Flux\ [10^{-7}\ erg\ s^{-1}\ cm^{-2}]$   & $\alpha$   & $\beta$    & $ \rm E_{ch}\ [keV]$    & Significance   \\\hline

 & BAT+GBM+KW & $33.60 \pm 5.03$  & $-0.60_{-0.06}^{+0.06}$ & $-1.99_{-0.09}^{+0.07}$ & $171.94_{-19.60}^{+22.56}$    & $\rm Stat/dof=1.22$    \\
  $93 - 102$ &XRT& $0.23 \pm 0.02$ &$-0.92_{-0.10}^{+0.10}$&//&//&$\rm Stat/dof=302.25/426$ \\
 &LAT&$0.03 \pm 0.02$&//&$-2.60 \pm 0.74$&//& $\rm TS =20$ \\ \hline
 & BAT+GBM+KW & $26.80 \pm 3.12$  & $-0.58_{-0.05}^{+0.05}$ & $-2.30_{-0.12}^{+0.10}$ & $204.54_{-19.69}^{+22.28}$    & $\rm Stat/dof=1.01$    \\
 $102 - 110$  &XRT& $0.27 \pm 0.03$ &$-0.92_{-0.11}^{+0.10}$&//&//&$\rm Stat/dof=332.48/505$ \\
 &LAT&$0.30 \pm 0.13$&//&$-5.36 \pm 1.68$&//&$\rm TS = 28 $ \\ \hline
 & BAT+GBM+KW & $14.60 \pm 1.27$   & $-0.60_{-0.04}^{+0.04}$ & $-3.09_{-0.42}^{+0.28}$ & $172.95_{-13.14}^{+15.21}$    & $\rm Stat/dof=1.16$   \\
 $110 - 118$&XRT&$0.28 \pm 0.03$  &$-0.85_{-0.10}^{+0.10}$&//&//&$\rm Stat/dof=384.71/450$ \\
 &LAT&$0.21 \pm 0.08$&//&$-3.81 \pm 0.99$&//&$\rm TS = 18$ \\ \hline
& BAT+GBM+KW  & $2.61 \pm 0.18$  & $-0.97_{-0.08}^{+0.04}$ & $-9.33_{-9.33}^{+19.33}$ & $160.40_{-19.59}^{+30.98}$    & $\rm Stat/dof=0.91$    \\
  $118 - 134$&XRT& $0.19 \pm 0.01$  &$-1.30_{-0.08}^{+0.08}$&//&//&$\rm Stat/dof=452.61/559$ \\
 &LAT&$0.25 \pm 0.13$ &//&$-3.62 \pm 1.22$&//&$\rm TS = 17 $ \\ \hline
&  BAT+GBM+KW  & $1.18 \pm 0.80$   & $-0.79_{-0.33}^{+0.39}$ & $-2.34_{-2.34}^{+0.41}$  & $78.06_{-30.80}^{+88.87}$     & $\rm Stat/dof=1.50$ \\
  $134 - 142$&XRT&$0.12 \pm 0.01$ &$-1.80_{-0.11}^{+0.11}$&//&//&$\rm Stat/dof=296.04/448$ \\
 &LAT&$0.19 \pm 0.09$&//&$-2.33 \pm 0.59$&//&$\rm TS = 24$ \\ \hline
&   BAT+GBM+KW  & $0.71 \pm 0.68$   & $-1.44_{-0.08}^{+0.20}$ & $-9.96_{-9.96}^{+19.96}$ & $540.28_{-307.37}^{+3048.23}$ & $\rm Stat/dof=0.93$\\
$142 - 159$ &XRT&$0.12 \pm 0.01$ &$-1.85_{-0.08}^{+0.08}$&//&//&$\rm Stat/dof=366.27/454 $ \\
 &LAT&$0.35 \pm 0.17$&//&$-3.39 \pm 0.99$&//&$\rm TS = 32$ \\ \hline
\hline
\end{tabular}
    \caption{Best-fit parameters obtained from the time-resolved independent spectral analysis of the XRT, BAT+GBM+KW and LAT datasets. We fit XRT and LAT data with a power law, while BAT+GBM+KW with Band model. We report the fluxes integrated over the respective instrument energy band, the low and high energy photon indexes $\alpha$ and $\beta$, respectively, and the Band characteristic energy $\rm E_{ch}$. Errors are reported with 1$\sigma$ confidence level.}
	\label{tab:timeres_fit_results}
\end{center}
\end{table*}

We perform a multi-instrumental spectral and temporal analysis, using the Heasarc package \textsc{XSPEC}\footnote{https://heasarc.gsfc.nasa.gov/xanadu/xspec/} and the Fermi Science tool \textsc{gtburst}\footnote{https://fermi.gsfc.nasa.gov/ssc/data/analysis/scitools/gtburst.html}. Details on data retrieving and reduction methods for all the instruments used in this work are reported in Appendix \ref{appendix:data}.\\
Initially, we divide the dataset in three blocks: XRT, LAT and BAT+GBM+KW data. For each block, we perform a time-resolved analysis, where the choice of the time-bins is driven by the time intervals where the \lat\ emission reaches a test statistics TS $>$ 10. XRT is fitted through \textsc{XSPEC} with a power law model, \texttt{powerlaw} in \textsc{XSPEC} notation, taking into account the Tuebingen-Boulder interstellar dust absorption ($\rm N_{\rm H} = 0.063 \times 10^{22} cm^{-2}$) and host galaxy dust absorption (source redshift z = 4.618) by using the \textsc{XSPEC} models \texttt{tbabs} and \texttt{ztbabs}, respectively. All the XRT spectra are consistent with zero intrinsic absorption. LAT spectra are also fitted with a power law through \textsc{gtburst}. BAT, GBM and KW spectra are jointly fitted through \textsc{XSPEC} with a Band model \citep{Band1993}, \texttt{grbm} in \textsc{XSPEC} notation, multiplied by cross-calibration constants. The results of these fits are reported in Table \ref{tab:timeres_fit_results}, while in Fig. \ref{fig:spectra+LCs} we show light-curves and photon indexes from the time-resolved spectral analysis.\\
From Fig.~\ref{fig:spectra+LCs}, lower panel, we notice that in the time-bins at t $> 118$ s both XRT and LAT spectra start to deviate from a single non-thermal component, i.e. the prompt emission spectrum. For this reason, we introduce two new time-bins, one between 94-118 s after the burst (hereafter \textit{prompt} time-bin) and the other between 118-241 s after the burst (hereafter \textit{prompt+afterglow} time-bin), and we perform the same analysis of the previous time-bins.\\
In the \textit{prompt} time-bin, XRT and LAT photon spectra can be described respectively by $\rm N_{XRT} \propto E_{XRT}^{-0.8}$ and $\rm N_{LAT} \propto E_{LAT}^{-4.4 \pm 0.9}$, both consistent with the low and high energy slopes inferred by fitting BAT+GBM+KW data. Conversely, in the \textit{prompt+afterglow} time-bin, we observe an excess in the spectrum at low and high energies, which leads to softer/harder slopes in XRT and LAT power laws (Fig.~\ref{fig:spectra+LCs}, upper panels). In particular, XRT and LAT spectra are best fitted by $\rm N_{XRT} \propto E_{XRT}^{-1.90}$ and $\rm N_{LAT} \propto E_{LAT}^{-2.1 \pm 0.3}$. This is indicative of a dominance of the keV-to-MeV (up to $\sim$ 0.25 GeV) prompt emission at early times, while at later times an additional component arises. We interpret this component to be afterglow emission from an external jet dissipation \citep{Paczynski1993,Meszaros1997}.\\

\section{Joint spectral analysis}\label{sec:joint_analysis}
The independent spectral analysis of three data blocks suggests the rise of a second emission component together with the prompt emission at times $\rm t > 118\ s$. In order to further investigate this scenario, we perform a joint spectral analysis by fitting the XRT, BAT, GBM, KW and LAT spectra through \textsc{XSPEC}. To take into account the differences among the instruments, we use cross-calibration constants and a mixed likelihood approach to weight correctly data errors (e.g. \citealt{ajello2020}), with different statistics depending on the instrument (PGstat for \gbm, Cstat for \lat, \xrt\ and Konus-\textit{Wind}, Gaussian statistics for \bat). \\
\subsection{Prompt emission from X-rays to high energies}\label{subsec:prompt_spectra}
\begin{figure}
\centering 
    \includegraphics[scale=0.6]{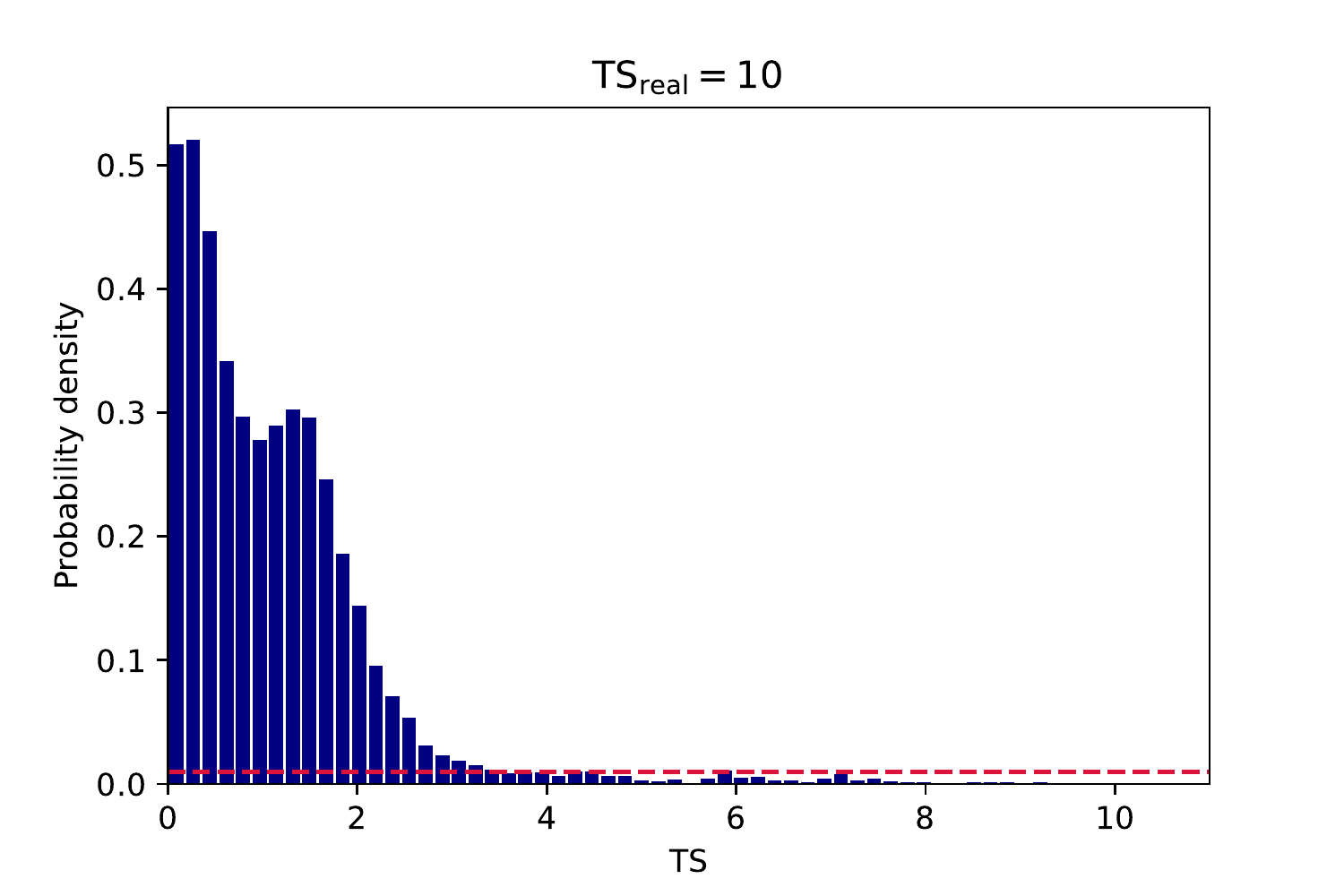}
    \caption{Distribution of simulated test statistics TS, obtained by fitting $10^3$ fake spectra during the \textit{prompt} time-bin (93-118 s) with Band (null model) and \textit{Band+cutoff} (alternative model). Fake spectra are produced by using as an input the Band model and its relative best-fit parameters obtained from the fit of the real data. The red-dashed line represents a probability density of $0.01$.}
	\label{fig:simulation_prompt}
\end{figure}

Initially, we fit the joint spectra in the \textit{prompt} time-bin with a Band function. We account for the galactic absorption, and we fix the extra-galactic absorption to zero, as evidenced by the previous independent analysis. We also compute the flux in the band $\rm 0.1\ keV-0.25\ GeV$ through the \textsc{XSPEC} model \texttt{cflux}, since the most energetic LAT photon has an energy of $\sim 250$ MeV in this particular time-bin.\\
We introduce an exponential cutoff at high energies, described in \textsc{XSPEC} by the model component \texttt{highecut}. The overall model (hereafter {\it Band+cutoff}) includes two new parameters: the energy  at which the cutoff starts to modify the base spectrum $\rm E_c$ and the energy which regulates the sharpness of the decay $\rm E_{fold}$. The sum of these two quantities provides the energy at which the flux drops by a factor $\rm 1/e$, namely $\rm E_{cutoff}=E_c + E_{fold}$. We test different $\rm E_{c}$ values and find consistent results on $\rm E_{cutoff}$, implying that the position of the $\rm E_{c}$ energy does not affect the main results of the analysis. Therefore, we fix $\rm E_c= 5\ MeV$ in order to minimise the number of free parameters.\\
To select which model fits better the data, we use the likelihood ratio test (LRT). 
We define $m_0$ the null model, in this case the simple Band model, and $m_1$ the alternative model, in this case the more complex \textit{Band+cutoff} model. We want to estimate if $m_1$ fits the data better than $m_0$ by computing the ratio between their mixed-likelihoods $L_0/L_1$. We define the test statistics ${\rm TS} = -2\log(L_0/L_1)$ and we use it as a best-fit indicator. If ${\rm TS}>0$ than $L_1 > L_0$, which means that the fit improves by using the model $m_1$ instead of $m_0$.\\
The data we used for the spectral analysis in the \textit{prompt} time-bin does not meet the regularity conditions required by Wilks' theorem (see \citealt{algeri2020}). Therefore, to assess if the improvement is statistically significant, we simulate $10^3$ spectra for each instrument employed in the joint spectral analysis, using as input the null model $m_0$ and its best-fit parameters obtained from the real data fit. Then, we fit all the simulated spectra with both $m_0$ and $m_1$ and we compute the relative test statistics $\rm TS_{sim}$. From the TS distribution we compute the probability density function (PDF) using the kernel density estimator. To reject the null hypothesis we require that the $p$-value, associated with the TS estimated from real data, is lower than a threshold $\alpha=0.01$.\\
From the real data fit, we measure $\rm TS_{real} = 10$, which corresponds to a $p$-value of $p = 2\times10^{-7}$, allowing us to reject the null hypothesis, implying that the addition of an exponential cutoff (\textit{Band+cutoff}) fits significantly better the spectral data with respect to the simple Band model. In Fig. \ref{fig:simulation_prompt} we show the probability density for different test statistics obtained by fitting $10^3$ fake spectra.\\
The best-fit parameters of the \textit{Band+cutoff} model are $\alpha = -0.78_{-0.02}^{+0.02}$, $\beta = -2.18_{-0.08}^{+0.06}$, $\rm E_{peak} = (316 \pm 20)\ keV$, $\rm E_{cutoff} = 85_{-26}^{+16}\ MeV$ and $\rm F_{0.1keV-0.25GeV} = (2.96 \pm 0.09) \times 10^{-6}\ erg\ s^{-1}\ cm^{-2}$.\\
We produce marginalized posterior distributions of the spectral parameters through the {\sc XSPEC} command {\tt chain} (See Appendix \ref{appendix:prompt_fit}).

\subsection{Modelling of the afterglow light-curve}\label{sec:afterglow}
\begin{figure}
\includegraphics[width=0.5\textwidth]{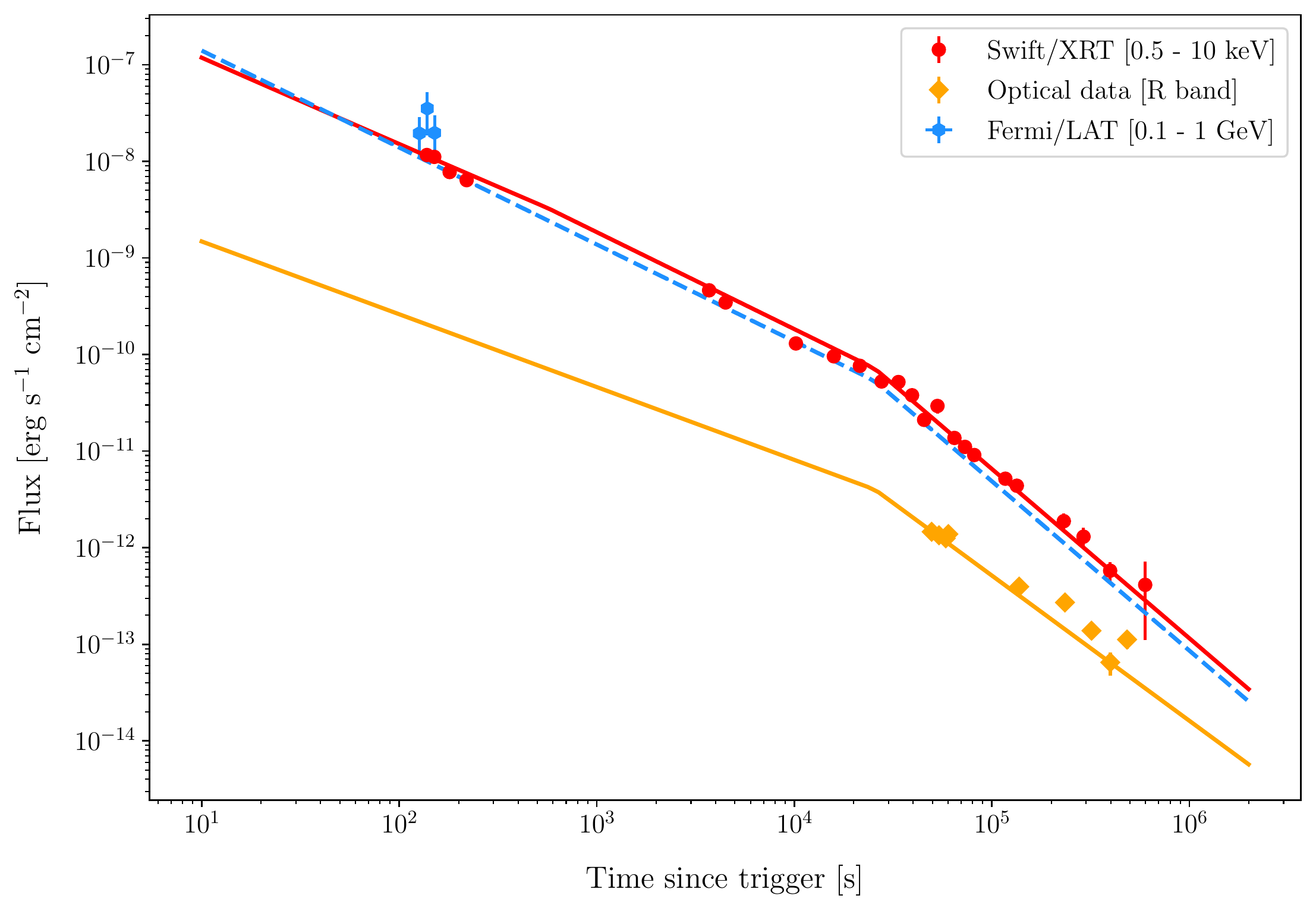}  
\caption{Optical (r-band, in orange), X-ray (0.5 - 10 keV, in red) and high energy (0.1 - 1 GeV, in blue) light-curves of GRB 220101A. Lines correspond to the afterglow model by the propagation of the forward shock in a homogeneous circum-burst medium. Early X-ray ($\rm <$ 134 s) and high-energy data ($\rm <$ 118 s) are not considered for the modelling of the joint light-curves.}
\label{fig:afterglow_LC}
\end{figure}

We use data provided by XRT (130 s - $\rm 6 \times 10^{5}$ s), LAT (118 - 160 s, three time-bins) and the r-filter optical data to infer the parameters of the external shock. First, we fit the X-ray light-curve empirically by smoothed broken power-law, which returns $\rm F_{X} \propto t^{-0.98 \pm 0.01}$ before the temporal break of $\rm t_{X} \approx 3 \times 10^{4}$ s and $\rm F_{X} \propto t^{-1.7 \pm 0.1}$ after. This temporal behaviour is in the best match with the forward shock propagating in the homogeneous circum-burst medium by requiring that the electron distribution index $\rm p>2$ \citep{Sari1998,Granot2002,Gao2013}. The temporal break in this scenario corresponds to the jet break, thus allowing us to constrain the opening angle of the jet once the density of the circum-burst medium and the kinetic energy of the jet are established. To get constraints on the jet opening angle, the density of the medium and the micro-physical parameters of the external shock, we model the joint optical-to-GeV light curve by the standard forward shock model in the homogeneous circum-burst medium. Since we observe the afterglow in the decaying phase (no peak is observed) and we deal with one of the most energetic GRBs (i.e. observed on-axis), we safely use the analytical expressions for the self-similar adiabatic solutions \citep{Granot2002,Gao2013}. We sample all the six parameters via MCMC, namely the isotropic equivalent kinetic energy of the jet $\rm E_{kin}$, the opening angle of the jet $\rm \theta_{j}$, the circum-burst medium density $\rm n$, electron distribution index $\rm p$, constant fraction of the shock energy that goes into electrons $\rm \epsilon_{e}$ and into the magnetic energy density $\rm \epsilon_{B}$. We include also one more parameter, the unknown absorption of the optical emission $\rm A_{R}$.\\
We explain the details of the MCMC in Appendix \ref{appendix:afterglow_fit}. The priors used for the MCMC analysis, together with the results of the fit, are reported in Table \ref{tab:mcmc_values}. Fig. \ref{fig:afterglow_LC}  shows the optical, X-ray and high energy light curves at $\rm t > 118\ s$ with the relative best-fit models.
\begin{table*}
	\label{table:mcmc}
	\begin{center}
		\begin{tabular}{c | c | c | c} 
			\hline\hline
			MCMC Parameter & Prior & Before LAT cut & After LAT cut  \\
			\hline
			$\rm \log E_{kin}\ [erg] $ & $(50, 60)$ & $56.54_{-1.01}^{+1.24}$ & $56.63_{-0.95}^{+1.03}$ \\[1.5ex]
			$\rm \log n\ [cm^{-3}] $ & $(-4, 4)$ & $1.58_{-2.66}^{+1.76}$ & $-1.11_{-1.98}^{+2.35}$ \\[1.5ex]
			$\rm \theta_j\ [deg]$ & $(0.1, 10)$ & $1.31_{-0.70}^{+1.08}$ & $0.63_{-0.24}^{+0.54}$ \\[1.5ex]
			$\rm p$ & $(2, 2.5)$ & $2.005_{-0.003}^{+0.006}$ & $2.006_{-0.003}^{+0.005}$ \\[1.5ex]
			$\rm \log \epsilon_e$ & $(-4, -0.5)$ & $-1.85^{+0.96}_{-1.31}$ & $-1.99_{-1.12}^{+0.94}$ \\[1.5ex]
			$\rm \log \epsilon_B$ & $(-8, -0.5)$ & $-4.38^{+1.83}_{-1.15}$ & $-2.57_{-1.75}^{+1.46}$ \\[1.5ex]
			$\rm A_R$& $(0, 3)$ & $2.05^{+0.09}_{-0.08}$ & $2.06_{-0.08}^{+0.10}$ \\[1.5ex]
			\hline
			Derived parameter &&& \\
			\hline
			$\rm \log E_{kin, \theta}\ [erg] $ &  & $52.93_{-0.93}^{+1.09}$ & $52.31_{-0.85}^{+1.33}$ \\[1.5ex]
			$\rm \log E_{\gamma, \theta}\ [erg] $ &  & $50.92_{-0.67}^{+0.52}$ & $50.29_{-0.42}^{+0.54}$ \\[1.5ex]			
			$\rm \eta_\gamma$ & & $ < 0.09 $ & $< 0.07$ \\[1.5ex]
			$\rm  \Gamma_{transp} $ & & $7554$ & $7090$ \\[1.5ex]
			$\rm  \Gamma_{LAT} $ &  & & $1270$ \\[1.5ex]			
			$\rm  \Gamma_{cut} $ & & & $1567$ \\[1.5ex]
			$\rm  \Gamma_{dec} $ & & & $578$ \\[1.5ex]
			\hline\hline

		\end{tabular}
	\end{center}
	\caption{Mean posterior values of the afterglow model parameters considering XRT and LAT data. We also report parameter values derived from afterglow MCMC analysis, namely the angle-corrected kinetic energy $\rm E_{kin, \theta}$, the angle-corrected energy $\rm E_{\gamma, \theta}$ emitted in the prompt phase, the relative energy conversion efficiency $\rm \eta_\gamma$ and upper/lower limits on $\Gamma_0$ ($\rm  \Gamma_{transp}$, $\rm  \Gamma_{LAT}$, $\rm  \Gamma_{cut}$, $\rm  \Gamma_{dec}$). All the estimates in this table are reported before and after cutting the posterior according to the LAT limit on coasting and deceleration afterglow phases. Errors and upper/lower limits are computed at 1$\sigma$ credible region.}
	\label{tab:mcmc_values}
\end{table*}

\section{Constraints on the prompt emission region}\label{sec:constraints_gamma}

\subsection{Limits from the afterglow emission}
The weakest lower limit on the bulk Lorentz factor can be obtained requiring that the highest energy of the photons observed during the afterglow emission can not exceed the maximum synchrotron frequency emitted by electrons in the jet comoving frame $\rm h\nu_{max} = hm_{e}c^{3}/2\pi^{2}e^{2}\ \simeq 22\ MeV$ \citep{Guilbert1983}. Given the maximum energy of the photon $\rm E_{LAT,max}=930\ MeV$ detected by \lat\ at 150 s, we place the lower limit on the bulk Lorentz factor of:
\begin{equation}\label{LF_burn}
\rm \Gamma_{0}> \Gamma_{burn-off} = \frac{E_{LAT,max}}{h\nu_{max}} (1+z) \simeq 237
\end{equation}
One should pay attention to this limit, since larger $\rm h\nu_{max}$ can be obtained in the shock accelerated electrons if the magnetic field is much stronger close to the shock front and decays downstream \citep{Kumar2012}. However, in our case the further lower limit on $\rm \Gamma_{0}$ exceeds $\rm \Gamma_{burn-off}$, therefore our general conclusions do not depend on the details of the magnetic field profile in the shock front.\\
Another lower limit on the bulk Lorentz factor can be found by the fact that we do not witness the peak of the afterglow emission \citep{Sari1999}. The upper limit on the peak time of the afterglow emission $\rm t_{p}<118$ returns then the lower limit on $\rm \Gamma_{0}$ \citep{Ghisellini2010,Ghirlanda2012,Lu2012,Nava2013}:

\begin{equation}\label{LF_aft}
\Gamma_{0}> \Gamma_{dec} =  k \left( \frac{E_{kin}}{n m_{p} c^{5}} \right)^{\frac{1}{8}} t_{p}^{-\frac{3}{8}} 
\end{equation}
Where $\rm k$ is the normalisation factor adopted from \citet{Nava2013}. 

\subsection{The compactness argument}

Constraints on the $\rm R_{\gamma}-\Gamma_{0}$ parameter space can be obtained from the compactness argument, which relies only on the prompt emission properties. In the comoving frame of the jet, high energy photons produce pairs. The optical depth to the pair production of a photon with an energy $\rm \epsilon^{\prime}$ (measured in the jet comoving frame) is defined as:
\begin{equation}\label{taugammagamma}
\tau_{\gamma \gamma}(\epsilon^{\prime})=\eta(\beta_e) \sigma_{T} \frac{U_{rad}^{\prime}(1/\epsilon^{\prime})}{m_{e}c^{2}} \delta R^{\prime}
\end{equation}
where $\rm U_{rad}^{\prime}$ is the comoving photon energy density, $\rm \delta R^{\prime}\approx R_\gamma/\Gamma_0$ is the comoving width of the jet shell, $\rm \eta(\beta_e)=(7/6) \cdot (2+\beta_e)^{-1}(1+\beta_e)^{-5/3}$ and $\rm \beta_e$ is the energy spectral index of the observed GRB spectrum \citep{Svensson1987}. To infer the $\rm R_{\gamma}-\Gamma_{0}$ relation for a given observable ($\rm E_{cut}$, spectral indices and the peak energy of the GRB spectrum $\rm E_{peak}$), one can either use the total radiated energy of the single pulse $\rm E_{rad}$ or its luminosity $\rm L_{rad}$ to define $\rm U_{rad}^{\prime}(1/\epsilon^{\prime})$. If $\rm E_{rad}$ is used, then $\rm \tau_{\gamma \gamma} \propto E_{rad}/R_{\gamma}^{2}$, while if $\rm L_{rad}$
is adopted the dependence is $\tau_{\gamma \gamma} \propto L_{rad}/(R_{\gamma} \Gamma_0^{2})$. Naturally, the difference in $\tau_{\gamma \gamma}$ for a $\gamma-$ray flash observed in the lab frame is factor of $\rm R_{\gamma}/c\Gamma_0^{2}$. However, we do use the spectrum in the \textit{prompt} time-bin to constrain the high energy spectral cutoff, therefore $\rm L_{iso}$ is the best proxi for $\rm L_{rad}$, while using $\rm E_{iso}$ instead of $\rm E_{rad}$ would overestimate the optical depth by the factor $T_{dur}c\Gamma_0^{2}/R_{\gamma}$, i.e. by 2-3 orders of magnitude, where $\rm T_{dur}$ is the time required for the spectral analysis. We notice that some works use $\rm E_{iso}$ as proxi for $\rm E_{rad}$ \citep{Gupta2008,Zhang2009}, while others adopt $\rm L_{iso}$ as proxi for $\rm L_{rad}$ \citep{Lithwick2001,Granot2002}. Alternatively, one can use $\rm E_{iso}$ as an approximation for $\rm E_{rad}$, but it requires a correction factor of $\rm t_{var}/T_{dur}$, where $\rm t_{var}$ is the variability time-scale \citep{Hascoet2012}.\\
Knowing that $\rm E_{cutoff} \simeq 80\ MeV$ in the observer reference frame, we can obtain an upper limit for $\Gamma_0$ by comparing $\rm E_{cutoff}$ in the observer and source reference frames:
\begin{equation}
    \Gamma_0 < \Gamma_{cut} = \frac{E_{cutoff}}{m_e c^2}(1+z)
\end{equation}
Given the measured isotropic equivalent luminosity $\rm L_{iso}$ between 94 and 118 s, the GRB peak energy $\rm E_{peak}$, energy spectral indices $\alpha_e$ and $\beta_e$ and the spectral cutoff $\rm E_{cutoff}$, by imposing $\rm \tau_{\gamma \gamma}=1$ we can derive a relation between $\rm R_\gamma$ and $\Gamma_0$ (\citealt{Ravasio_thesis2022}, article in preparation):
\begin{equation}\label{eq:R-G relation}
\begin{split}
R_\gamma = & \eta(\beta_e) \sigma_T \frac{L_{iso}}{4\pi} \frac{(1-\alpha_e)(\beta_e-1)}{\beta_e-\alpha_e} \left(\frac{(m_e c^2)^2}{E_{peak}}\right)^{-\beta_e} \times \\
& \times \frac{E_{cutoff}^{\beta_e}}{c E_{peak}}\ {\Gamma_0}^{-2-2\beta_e}   
\end{split}
\end{equation}
In this equation, both $\rm E_{peak}$ and $\rm E_{cutoff}$ are corrected for the redshift.

\subsection{Upper limit on $\rm  \Gamma_{0}$ from the fireball dynamics}
In the hot fireball model, by requiring that the GRB production site is above the jet photosphere, i.e. $\rm R_{\gamma}>R_{ph}(\Gamma_{0})$, one can obtain the following upper limit on $\rm \Gamma_{0}$:
\begin{equation}\label{taugammagamma}
\Gamma_{0}< \Gamma_{transp} =\left( \frac{L_{iso}\sigma_{T}}{8 \pi m_{p} c^{3} \eta_{\gamma} r_{0}} \right)^{1/4}
\end{equation}
where $\rm \eta_{\gamma}=E_{iso}/(E_{iso}+E_{kin})$ is the efficiency of the prompt emission production and $\rm r_{0} \sim 10$ km is the initial fireball radius which is of order of the central engine one \citep{daigne2002}.\\

\begin{figure*}
\centering 
\includegraphics[width=0.9\textwidth]{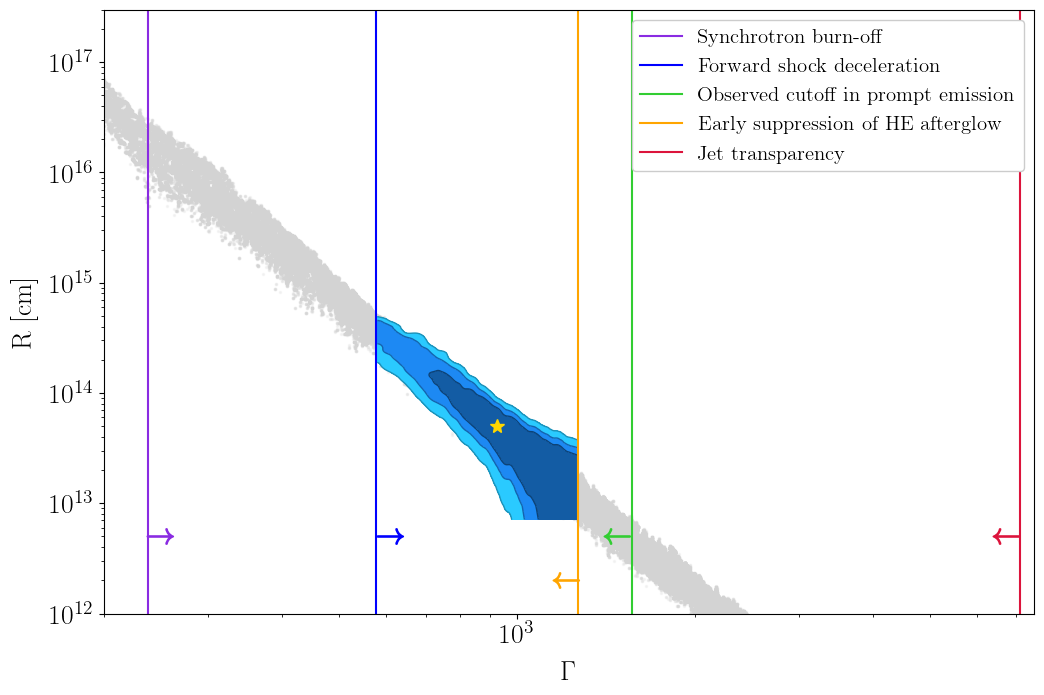}  
\caption{Constraints on the $\rm R_{\gamma}-\Gamma_{0}$ prompt emission region. The grey points are drawn from the $\rm R_{\gamma}-\Gamma_{0}$ distribution after considering only the condition on the cutoff (first scenario), corresponding to the allowed parameter space derived from the opacity argument (requiring $\rm \tau_{\gamma \gamma}=1$ at $\rm E_{cutoff}$). We show the lower limits on $\rm \Gamma_{0}$ from the maximum synchrotron emitted energy (purple line) and from external shock deceleration (blue line) and the upper limits on $\rm \Gamma_{0}$ from cutoff observation in prompt emission (green line), GeV afterglow emission at coasting and deceleration phase (orange line) and jet transparency requirement (red line). These upper/lower limits are reported with 1$\sigma$ confidence level. The blue-shaded areas correspond to the $R_{\gamma}-\Gamma_{0}$ parameter contours at 1$\sigma$, 2$\sigma$ and 3$\sigma$ confidence levels. The contour plot and relative 1$\sigma$ estimate of $\rm R_\gamma$ and $\Gamma_0$ (in yellow) are both obtained in the last scenario, which takes into account all the conditions on $R_{\gamma}-\Gamma_{0}$ parameters.}
	\label{fig:R-Gamma_plane}
\end{figure*}

\subsection{Upper limit on $\rm  \Gamma_{0}$ from the high energy afterglow emission}

The early rise of the forward shock emission at high energies, quantified by the observed fluence $\rm S_{HE, rise}$ in the energy band 0.1 - 1 GeV, strongly depends on the bulk Lorentz factor as $\rm S_{HE, rise} \propto \Gamma_0^{2p+4}$.
Similarly, before the jet starts to decelerate, at the peak of the afterglow emission, the observed high energy afterglow fluence $\rm S_{HE, peak}$ depends on the observed prompt fluence $\rm S_{iso}$ and on its efficiency $\rm \eta_\gamma$ as $\rm S_{HE, peak} \propto \frac{1-\eta_{\gamma}}{\eta_{\gamma}}S_{iso}$ \citep{Nava2017} . These relations are based on two assumptions: first, the high energy afterglow emission corresponds to the synchrotron emission above the cooling frequency $\rm \nu_{c}$ \citep{Kumar2010} and, second, the inverse Compton (IC) cooling of electrons above $\rm \nu_{c}$ is above the Klein-Nishina limit for the typical parameters of the external shock of GRBs \citep{Beniamini2015}.\\
We observe that the sub-GeV emission at early times is dominated by prompt emission (Fig.\ref{fig:spectra+LCs}, upper-left panel). In addition, from the modelling of the high energy afterglow light curve (Fig.\ref{fig:afterglow_LC}), we infer that the coasting and deceleration phases take place before $\sim 100$ s from the burst, simultaneously with prompt emission. This implies that, in the prompt time-bin (94 - 118 s), high energy afterglow fluence at coasting and deceleration phases can not exceed the observed high energy prompt emission $\rm S_{HE, prompt}$ in the same time-bin. Therefore, by requiring that both $\rm S_{HE, rise}\ and\ S_{HE, peak}$ are smaller than $\rm S_{HE, prompt}$ , we can not only find a new constraint for $\rm \Gamma_0 < \Gamma_{LAT}$, but also perform a cut on the posterior distributions obtained from the afterglow light curve modelling. The new posterior distributions lead to more precise afterglow parameter estimates, thus improving the other upper/lower limits on $\rm \Gamma_0$.\\
To take into account the additional electron cooling by the Inverse Compton, we assume fast cooling mode of the synchrotron radiation, i.e. $\rm Y \approx (\epsilon_{e}/\epsilon_{B})^{1/2}$ and we correct it for the Klein-Nishina effect \citep{Nakar2009}.\\
One could think of an alternative scenario, where the GeV afterglow emission is absorbed by the MeV prompt emission produced below the afterglow deceleration radius \citep{Zou2011}. In that case, the fact that we do not observe bright GeV afterglow emission is due to the prior MeV-GeV prompt-to-the-afterglow photons absorption. However, in that case we would require the total energy of the prompt emission to be much more than what we have observed, since we observe a high energy power-law in the prompt emission spectrum. Dealing with one of the most energetic GRB, we disfavor this scenario.\\

\subsection{$\rm R_{\gamma}-\Gamma_{0}$ constraints}
We develop a method to constrain the allowed parameter region in the $\rm R_{\gamma}-\Gamma_{0}$ plane for GRB 220101A. The method consists in building a $\rm \Gamma_0$ parameter distribution based on the previous prompt and afterglow analyses (Sec.~\ref{sec:joint_analysis}), together with the $\Gamma$ limits discussed in Sec.~\ref{sec:constraints_gamma}.\\
By randomly sampling $N=8\times 10^5$ times the posterior distributions of prompt spectral parameters ($\rm F$, $\alpha$, $\beta$, $\rm E_{ch}$, $\rm E_{cutoff}$) and of the afterglow parameters ($E_{kin}$, $\theta_j$, $n$, $p$, $\rm A_r$, $\epsilon_e$, $\epsilon_B$) previously obtained (see Appendix \ref{appendix:mcmc} for details), we can build distributions also for the $\Gamma$ limits ($\rm \Gamma_{transp}$, $\rm \Gamma_{LAT}$, $\rm \Gamma_{cut}$, $\rm \Gamma_{dec}$).
Once the conditions for $\Gamma_0$ are obtained, we impose its distribution to be uniform and restricted by the $\Gamma$ limits. This new distribution, together with the spectral parameter ones, are sampled again in order to obtain $N$ estimates of $R_\gamma$, by requiring that $\rm \tau_{\gamma\gamma} = 1\ at\ E_{cutoff}$ (Eq.~\ref{eq:R-G relation}). At the end of these steps, we can define two distributions for $\rm R_{\gamma}$ and $\Gamma_{0}$, whose values will occupy a given restricted region of the parameter space depending on the conditions imposed on $\Gamma_0$. The more constraints we use, the smaller this region is, hence the more precise is the estimate of these values. Therefore, we present three scenarios where we take into consideration different conditions, in order to highlight the role of each observational feature. We report constraints, upper and lower limits with 1$\sigma$ confidence level, i.e. computing the 50-, 84- and 16-percentile of the parameter distribution, respectively.\\
In the first scenario, we only consider the condition on fireball dynamics, providing a wide flat distribution for $\rm \Gamma_0$ in the range between 1 and $\rm \Gamma_{transp}$. This returns $\rm logR_\gamma[cm] > 10\ $ and $\rm \Gamma_0 < 6343$.\\
In the second scenario, we add the condition derived from LAT observation in both prompt and afterglow emissions. We take $\rm \Gamma_{burn-off}$ as lower limit for $\rm \Gamma_0$ and the minimum between $\rm \Gamma_{cut}$, $\rm \Gamma_{LAT}$ and $\rm \Gamma_{transp}$ as upper limit. We obtain $\rm logR_\gamma[cm] = 14.09_{-0.73}^{+1.22}$ and $\rm 400<\Gamma_0<1100$.\\
We compute upper/lower limits in this and in the following steps after performing a cut in the afterglow posterior distributions. We define an initial $\rm \Gamma_0$ uniformly distributed between $\rm \Gamma_{burn-off}$ and the $\rm \Gamma_{transp}$ estimate from the previous step. Afterwards, we sample the initial posteriors, and accept the values of $\rm \Gamma_0$ and prompt/afterglow parameters that satisfy the conditions $\rm S_{HE, rise} <  S_{HE, prompt}$ and $\rm S_{HE, peak} <  S_{HE, prompt}$. We define $\rm \Gamma_{LAT}$ as the maximum value of the $\rm \Gamma_0$ distribution after LAT cut. We report the 1$\sigma$ values of afterglow and derived parameters after the posterior cut in Table~\ref{tab:mcmc_values}.\\
In the third scenario, we add the condition on the deceleration phase in the X-ray afterglow observed by \xrt. Therefore, we define the maximum between $\rm \Gamma_{dec}$ and $\rm \Gamma_{burn-off}$ as lower limit on $\rm \Gamma_0$ and the minimum between $\rm \Gamma_{LAT}$, $\rm \Gamma_{transp}$ and $\rm \Gamma_{cut}$ as upper limit. The $\Gamma$ limits posteriors are considered after LAT cut. We obtain $\rm logR_\gamma[cm] = 13.7_{-0.4}^{+0.6}$ and $\rm 700<\Gamma_0<1160 $.\\
In Fig.~\ref{fig:R-Gamma_plane} we show how the initial parameter space is reduced after taking into consideration all the conditions on $\Gamma_0$ . We notice that the most stringent conditions are the ones obtained by forward shock deceleration from X-ray afterglow light curve and by the high energy spectral cutoff during prompt emission. 


\section{Discussion and Conclusions}
Given the values of $\rm R_{\gamma} \approx 4.5 \times 10^{13}$ cm and the median $\rm \Gamma_0 \approx 900$, we can provide an estimate of the magnetic field $\rm B^{\prime}$ (in the comoving frame) that matches the observed peak energy of the GRB spectrum $\rm E_{peak} \approx 300$ keV, assuming that the dominant radiative process is synchrotron emission. In addition, we assume $\nu_c \simeq \nu_m$, since the low energy spectral index $\alpha$ is roughly consistent with the usual value of -2/3 and we do not observe any additional spectral break. If we consider the electrons to produce the observed spectrum, we require:
\begin{equation}
\rm B_{e}^{\prime} \approx (190 \, G) \,  R_{13.65}^{-2/3}\ \Gamma_{0,900} \, h\nu_{obs,300}^{-1/3}  
\end{equation}
In this estimate we consider that the synchrotron cooling time-scale coincides with the angular and radial time-scales, i.e. $\rm t_{syn}=t_{R}=t_{\theta}=R/2c\Gamma_0^{2}$.\\
For such low values of the magnetic field and the luminosity observed, the radiation energy density in the emitting region could imply a non-negligible synchrotron-self-compton (SSC) cooling of the particles \citep{Ghisellini2020}. 
An electron-synchrotron component driven by a small magnetic field $\rm B_{e}^{\prime} \approx 190 \, G$ in a relatively compact emitting region ($\rm R_{\gamma} \approx 4.5 \times 10^{13}$ cm) would be easily overshined by the SSC emission (of at least a factor $10^4$), not matching the spectral behaviour observed in this source \citep{Kumar2008,Beniamini2013,Oganesyan2019,Ghisellini2020}.\\
This tension can be alleviated if one considers protons as synchrotron emitters in prompt emission. In fact, protons can naturally produce marginally fast cooling synchrotron spectra, allowing for large magnetic fields of the order of:
\begin{equation}
\rm B_{p}^{\prime} \approx (5 \times 10^{7} \, G) \, R_{13.65}^{-2/3}\ \Gamma_{0,900} \, h\nu_{obs,300}^{-1/3}
\end{equation}
Which would require a high (collimation-corrected) Poynting flux (see \citealt{Florou2021}):
\begin{equation}
 \rm P_{B,p} \approx (5 \times 10^{54}\ erg/s)\ \theta_{j,1deg}^{2}\ B_{7.72}^{2}\ \Gamma_{0,900}^{2} R^2_{13.65}   
\end{equation}
In the estimates mentioned above we have assumed that the GRB spectrum arises from marginally fast cooling electrons/protons. If we relax this requirement, i.e. assume that $\rm t_{syn} \ge R/2c\Gamma_0^{2}$, then our estimates on $\rm B^{\prime}$ would be only upper limits.\\
Nonetheless, it is expected that, in the assumption of electrons and protons injected with the same Lorenz factor distribution, the electron-synchrotron component luminosity would be smaller by a factor $m_p/m_e$, allowing us to neglect this contribution \citep{Ghisellini2020}.\\
This scenario does not include the contribution that electrons can have in the overall spectrum through synchrotron cooling, and possibly SSC and inverse Compton (IC) with the proton-synchrotron photons. Deeper investigations of the role of electrons in the proton-synchrotron scenario are required to assess its capability of explaining current observations (see \citealt{Florou2021,Begue2021}).
We also notice that the observed luminosity of GRB 220101A together with the constrained values of $\rm R_{\gamma}$ and $\rm \Gamma_{0}$ suggest that the proton-synchrotron emission dominates over the synchrotron emission from the Bethe-Heitler pairs \citep{Begue2021}.\\
Moreover, it is consistent with the low energy spectral shape of the GRB 220101A, $\rm \alpha=0.70$ and the adiabatic cooling inferred from the X-ray decline of the prompt emission pulses \citep{Ronchini2021}. In this scenario, the GRB emitting particles (protons) do not cool efficiently in a dynamical time-scale and the GRB variability is given purely by the adiabatic expansion time, i.e. $\rm R/2c\Gamma_0^{2}$. Clearly, this corresponds to very low prompt emission efficiency, leaving most of the energy to dissipate in the afterglow phase. However, electrons will loose all their energy given large magnetic fields, producing both high and very high energy (VHE) emissions. In this scenario, what we observe as GRB at keV-MeV range is only adiabatically cooling proton emission. \citet{Ghisellini2020} showed that if electrons and protons have the same Lorentz factor distribution, then we would expect an emission component at $\rm h\nu_{obs} \approx 0.5 \, GeV \, h\nu_{obs,300}$ with the luminosity of $\rm L_{iso} \approx 2 \times 10^{50}$ erg/s. In the case of electrons and protons sharing the same energy distribution, we would expect the electron-synchrotron component to peak at $\rm 2 \times 10^{15}$ eV with the same proton-synchrotron luminosity. In the latter case, one should carefully take into account for the pair cascade caused by these extremely high-energy photons. Nevertheless, the observations of the GRB prompt emission spectra at high and very high energies could be a powerful tool to discriminate the GRB emission models and to constrain the acceleration processes by identifying the relative proton-to-electron energy ratio.\\
We want to stress the fact that the conclusions of this work are drawn within the framework of the synchrotron model. We do not discuss the implication of a sub-photospheric emission \citep{Rees2005, Peer2008}, magnetic reconnection in a highly magnetized ejecta \citep{Zhang2011}, or hybrid jets \citep{gao&zhang2015}. Despite this, it is worth to mention that the most stringent constraints we find in the $R_\gamma-\Gamma$ plane are driven from observations, not requiring any prior assumption on the prompt physics.\\
In this work we analyse one of the most energetic GRB ever observed. 
Estimated from the KW detection (GCN 31433), the burst ${\rm E_{iso}}$ is $3.64_{-0.22}^{+0.25} \times 10^{54}$~erg, which is within the highest $\sim$2\% for the KW sample of 338 GRBs with known redshifts \citep{Tsvetkova2017, Tsvetkova2021}.
With this ${\rm E_{iso}}$ and the rest-frame peak energy of $1416_{-157}^{+152}$~keV, GRB 220101A is within 68\% prediction bands for Amati relation for the same sample of long KW GRBs with known redshifts.\\
The joint spectral and temporal analysis of the source shows the presence of the afterglow emission in the X-ray and high energy bands from $\rm 118\ s$. The measured cutoff energy $\rm E_{cutoff} = 85_{-26}^{+16}$ MeV and minimum afterglow observations revealed to be powerful tools to constrain the dynamics and dimension of the prompt emitting region, leading to the stringent constraints on $\rm logR_\gamma[cm] = 13.7_{-0.4}^{+0.6}$ and $\rm 700<\Gamma_0<1160 $, in favor of a proton-synchrotron scenario rather than an electron-synchrotron one. The inferred radius of the prompt emission is above the  jet photosphere and below the typical magnetic reconnection regions. \\
More observations in the MeV-GeV and very high energy domains are necessary to fully uncover the physics of the GRB jet dissipation and acceleration processes. 

\begin{acknowledgements}
The authors thank A. Celotti, O. S. Salafia, G. Ghirlanda and G. Ghisellini for fruitful discussions. GO and MB acknowledge funding from the European Union’s Horizon 2020 Programme under the AHEAD2020 project (grant agreement n. 871158). BB and MB acknowledge financial support from MIUR (PRIN 2017 grant 20179ZF5KS). AT and DF acknowledges support from RSF grant 21-12-00250. G.O. and M.B. acknowledge financial contribution from the agreement ASI- INAF n.2017-14-H.0. This work made use of data supplied by the UK Swift Science Data Centre at the University of Leicester.
\end{acknowledgements}

\appendix


\section{Data}\label{appendix:data}
\subsection{Swift/XRT}
We have downloaded the data provided by the X-Ray Telescope (0.5 - 10 keV, XRT) on-board the Neil Gehrels Swift Observatory ({\it Swift}) from the {\it Swift} Science Data Center supported by the University of Leicester \citep{Evans2009}. Nine time-bins (93 - 3788 s from the GRB trigger time) in the Window Timing mode and  18 time-bins ($\rm 4 \times 10^{3} - 6 \times 10^{5}$ s) in the Photon Counting mode were selected for the time-resolved spectral analysis to evaluate the temporal and spectral evolution of the X-ray emission during the prompt and the afterglow phases. Additional spectra at the early times are retrieved to perform joint \gbm, \ \lat, \bat\ and Konus-{\it Wind} analysis. The choice of the time-intervals was driven by the significant \lat\ detection. We adopt Cash statistic to fit XRT spectra.

\subsection{Swift/BAT}
The data from the Burst Alert Telescope (15 - 150 keV, BAT) were downloaded from the {\it Swift} data archive. The FTOOLS {\tt batmaskwtevt} and {\tt batbinevt} pipelines are used to extract the background-subtracted mask-weighted BAT light-curves. To produce BAT spectra and the corresponding response files, we have used the {\tt batbinevt} task together with {\tt batupdatephakw}, {\tt batphasyserr} and {\tt batdrmgen} tools. We have applied Gaussian statistics to fit the BAT spectra. 

\subsection{Konus-WIND}
The Konus-Wind instrument (KW; \citealt{aptekar1995}) is a $\gamma$-ray spectrometer consisting of two identical NaI(Tl) detectors, S1 and S2, which observe the southern and northern ecliptic hemispheres, respectively.
Each detector has an effective area of 80--160~cm$^2$, depending on the photon energy and incident angle, and collects the data in $\sim$20~keV--20~MeV energy range.
GRB~220101A triggered the S2 detector of the KW at T0(KW)=05:11:35.828 UT.
For this burst, the triggered mode light curves are available, starting from T0(KW)-0.512~s, in three energy windows G1($\sim$20--80~keV), G2($\sim$80--330~keV), and G3($\sim$330--1320~keV), with time resolution varying from 2~ms up to 256~ms and the total record duration of $\sim$230~s.
The burst spectral data are available, starting from T0(KW), in two overlapping energy intervals, PHA1 (20--1300~keV) and PHA2 (270~keV--16~MeV).
The total duration of the spectral measurements is $\sim$490~s.
The KW background is very stable and assumed to be at constant level during the triggered mode record.
For GRB~220101A \citep{tsvetkova2022, tsvetkova2022corr}, we constructed the background spectrum as a sum of spectra outside the burst emission episodes,
from $\sim$T0(KW)+205~s to $\sim$T0(KW)+435~s.
With $>$100 counts per energy channel, the background is assumed to be Gaussian, and the errors are computed as a square root of the channel counts.
When fitting the KW data, the $\chi^2$ statistics is typically applied to the spectra grouped to have min 20 cnts per energy bin, or pgstat can be used with the data grouped to min 1 cnt/bin.
In the latter case, cstat can also be used, which yields nearly the same results as pgstat.
A more detailed description of the KW data and the data reduction procedures can be found, e.g., in \citet{Svinkin2016, Tsvetkova2017, Tsvetkova2021}.

\subsection{Fermi/GBM}
We have selected two sodium iodide (NaI, 8-900 keV) detectors, namely NaI-6 and NaI-7, and one bismuth germanate (BGO, 0.3-40 MeV) detector BGO-1 to retrieve the data from the Gamma-ray Burst Monitor (GBM) on-board the Fermi Gamma-ray Space Telescope ({\it Fermi}). \gbm\ data are extracted by the {\sc gtburst} tool. We have excluded the energy bins below 8 keV and above 900 keV for NaI detectors and below 300 keV and above 10 MeV for the BGO-1 detector. To fit the \gbm\ spectra, we have applied PGSTAT likelihood.    

\subsection{Fermi/LAT}
The Large Area Telescope (LAT) on board \textit{Fermi} is sensitive to the gamma-ray photons of energy between 30 MeV and 300 GeV \citep{2013ApJS..209...34A}. 
We use {\sc gtburst} tool to extract and analyse the data. 
The source (R.A. $= 1.35^{\circ}$ and Dec. $= 31.77^{\circ}$) was inside the field of view (FoV) of LAT until around 1400\,s after the trigger.
For this analysis, we use a region of interest (ROI) of 12$^{\circ}$ centred at the burst position provided by {\it Swift}/BAT \citep{2022GCN.31347....1T}. 
As the spectral model, particle background and the Galactic component we assume "{\tt powerlaw2}", "{\tt isotr template}" and "{\tt template (fixed norm.)}" respectively.
The estimation of flux in the energy between 100\,MeV to 10\,GeV is performed with the "{\tt unbinned likelihood analysis}". Due to the low statistics in the \lat\ temporal bins, we can not perform a binned likelihood analysis. The highest energy of the photon associated with GRB 220101A has energy of 930 MeV at 150 s from the GRB trigger time. Most of the photons have energies between 100 and 250 MeV and they are detected during the main prompt emission episode (observed by BAT, GBM and KW). The time-bins for the joint spectral analysis were chosen requiring significant \lat\ detections (minimum test statistics $\rm TS >10$). No LAT LLE data are available for this burst. We generate LAT count spectrum thorugh {\sc gtburst} using the Standard ScienceTool\footnote{https://fermi.gsfc.nasa.gov/ssc/} pipeline {\tt gtbin}. In addition, we produce the background counts and response files using {\tt gtbkg} and {\tt gtrspgen}, respectively (e.g. \citealt{ajello2020}). We fit the LAT spectrum on {\sc XSPEC} using Cash statistics.

\subsection{Optical data}
GRB 220101A has been followed up by numerous optical telescopes. We have selected the r-band observations (AB system) from the GCN Circulars Archive to use single-filter homogeneous optical data for the afterglow modelling. We ignore the early optical detection by {\it Swift}/UVOT at $\rm \sim$ 150 s \citep{GCN31351}, since the single bright optical detection is not informative enough to discriminate between the forward and reverse shock contributions. We include in the analysis data from the Liverpool telescope \citep{GCN31357, GCN31425}, the Tautenburg 1.34m Schmidt telescope \citep{GCN31401}, the CAFOS instrument \citep{GCN31388} and the Konkoly Observatory \citep{GCN31361}. 

\section{Monte Carlo Markov Chain}\label{appendix:mcmc}

\subsection{Prompt fit}\label{appendix:prompt_fit}
\begin{figure*}
\centering 
\includegraphics[width=0.7\textwidth]{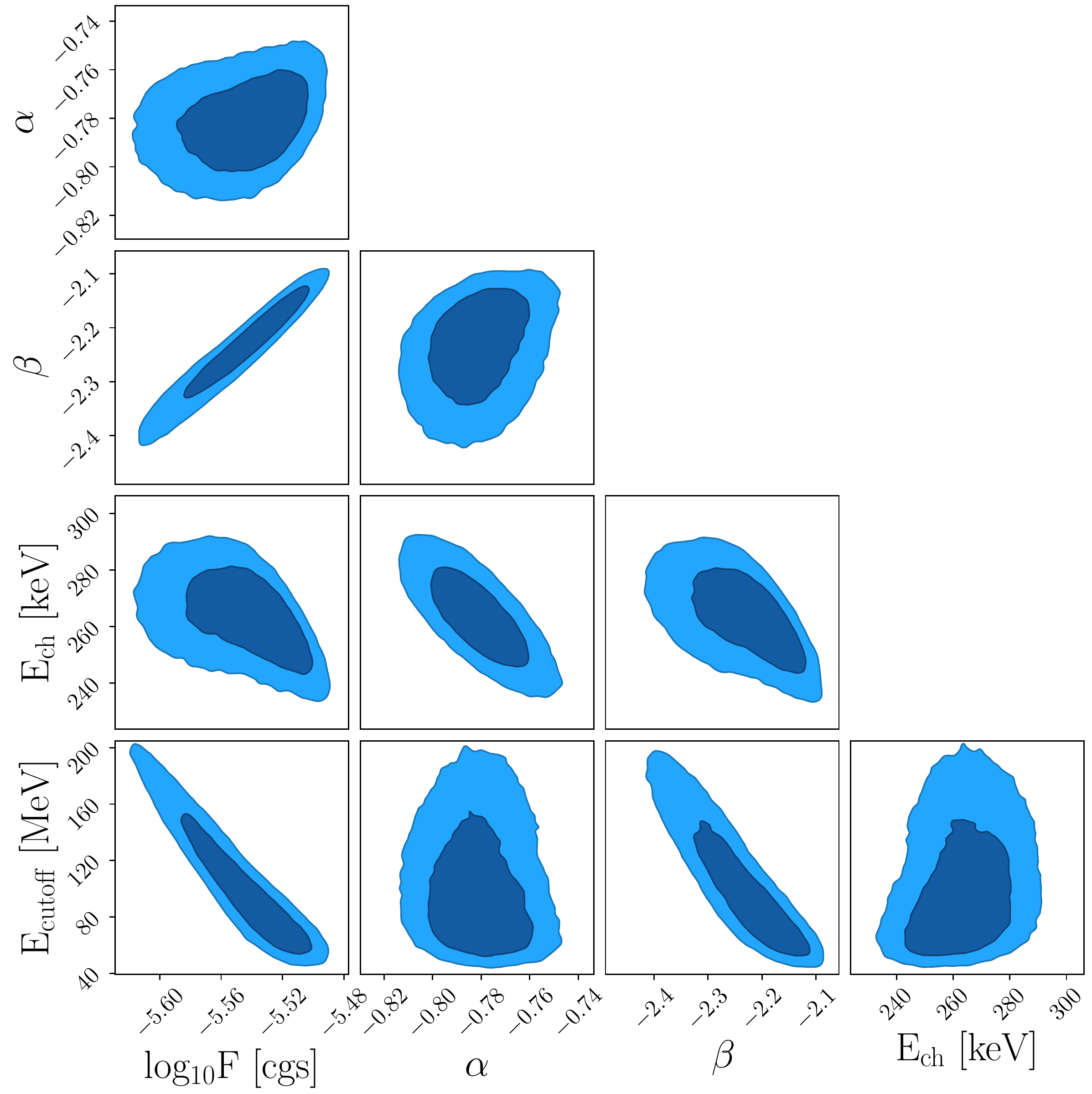}  
\caption{Corner plot showing the parameter contours obtained from the MCMC of the prompt fit parameters, namely the low and high energy slopes $\alpha$ and $\beta$, respectively, the characteristic energy $\rm E_{ch}$, the cutoff energy $\rm E_{cutoff}$ and the flux integrated between 0.1 keV-0.25 GeV. The 1$\sigma$ contour region is shown in dark blue, while in light blue we show the 2$\sigma$ contour region.}
	\label{fig:chain_corner}
\end{figure*}

We performed a joint spectral fit of XRT+BAT+GBM+KW+LAT data during the \textit{prompt} time-bin (94-118 s), which is described by a Band function with a cutoff at $\rm E_{cutoff} \simeq 80\ MeV$. For the \textit{Band+cutoff} model, we performed a Monte Carlo Markov Chain (MCMC) to sample the posterior distribution of the fitted parameters, using the \textsc{XSPEC} task \texttt{chain}.\\
This analysis returns, for each model parameter, a chain of parameter values whose density gives the parameter probability distribution. We employ the Goodman-Weare algorithm, requiring $N_{walkers} = 4$ walkers and $N_{iter} = 10^6$ iterations. Since the starting parameters are far from convergence, we ignore the first $N_{burn} = 5\times10^4$ steps. The walkers are initialised by drawing from a multi-Normal distribution whose variance matrix is based on the covariance matrix obtained from the previous \textsc{XSPEC} fit. The parameter contours obtained from the prompt MCMC is shown in Fig.~\ref{fig:chain_corner}.\\
\begin{figure*}
\centering 
\includegraphics[width=0.7\textwidth]{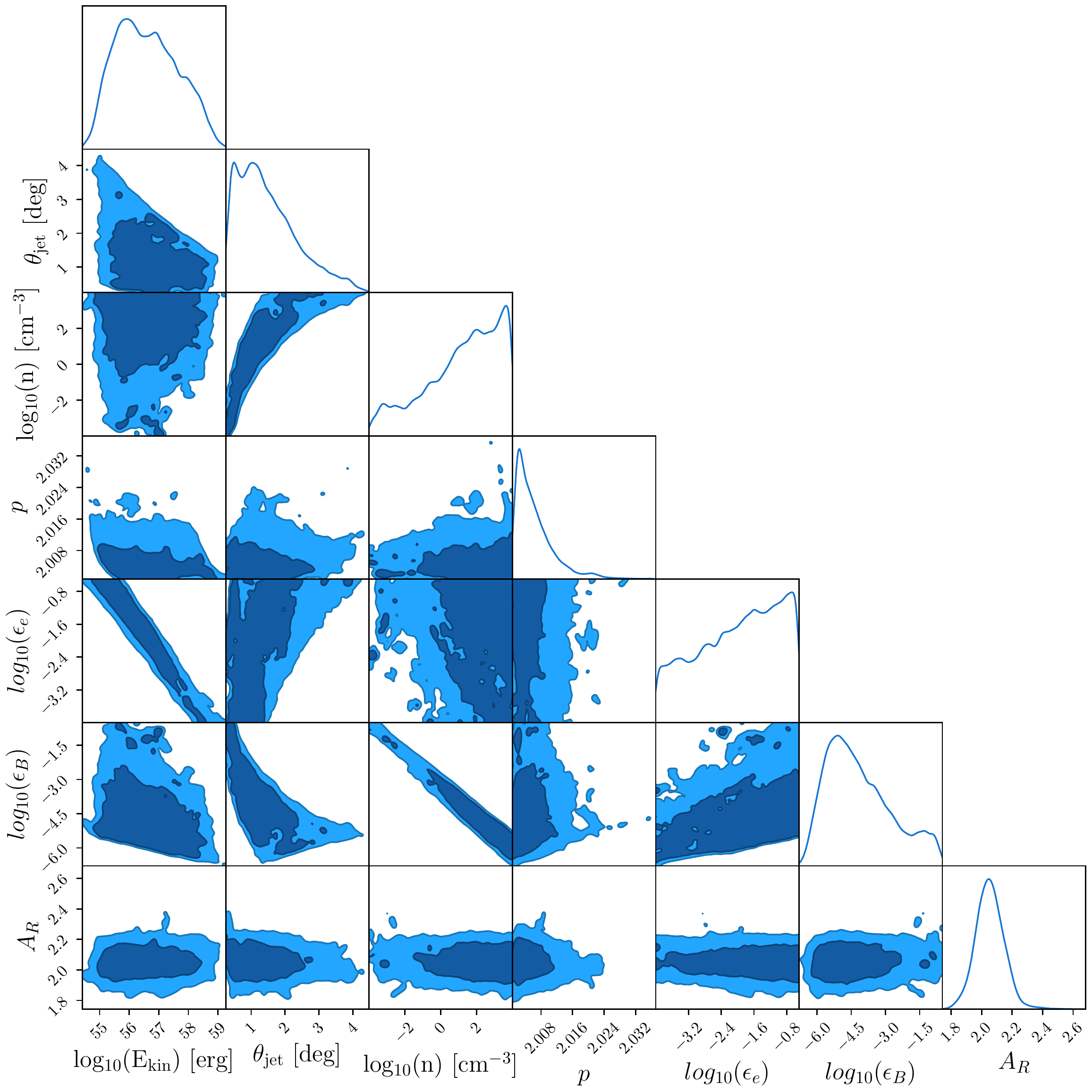}  
\caption{Corner plot showing the parameter marginalized posterior distributions and contours from the MCMC performed for the afterglow analysis. The best-fit values and priors are shown in Table \ref{tab:mcmc_values}. The parameters fitted in the MCMC are the isotropic equivalent energy $E_{kin}$, the jet opening angle $\theta_j$, the circum-burst medium density $n$, the electron distribution index $p$, absorption of optical emission $\rm A_r$, constant fraction of the shock energy that goes to electrons $\epsilon_e$ and into magnetic energy density $\epsilon_B$.}
	\label{fig:mcmc_corner}
\end{figure*}

\subsection{Afterglow fit}\label{appendix:afterglow_fit}
We fit 6 afterglow parameters using a MCMC approach, namely the isotropic equivalent kinetic energy of the jet $\rm E_{kin}$, the opening angle of the jet $\rm \theta_{j}$, the circum-burst medium density $\rm n$, electron distribution index $\rm p$, constant fraction of the shock energy that goes into electrons $\rm \epsilon_{e}$ and into the magnetic energy density $\rm \epsilon_{B}$ and the absorption of the optical emission $\rm A_{R}$.\\
Our observables are the flux density estimates measured in the optical, X- and $\gamma$-ray band $\rm F_{\nu, i}$ and the photon indexes in the X- and $\gamma$-ray band $\alpha_i$. Each observable $\rm \theta_i \in \{ F_{i}, \alpha_i \}$ contributes to the overall log-likelihood with an additive term, given by:
\begin{equation}
     \log L_i = -\dfrac{1}{2}\ \dfrac{(\theta_{m,i} - \theta_i)^2}{\sigma_{\theta,i}^2} -\dfrac{1}{2} \ln(\sigma_{\theta,i}^2)
\end{equation}
Where $\rm \theta_{m,i}$ is the observable predicted by the model and $\sigma_{\theta,i}$ is the associated uncertainty. Since GRB 220101A is particularly luminous and we do only observe the decaying phase of the afterglow, it is safe to assume it is on-axis. Therefore, we employ an analytical model based on self-similar adiabatic afterglow solutions \citep{Granot2002, Gao2013} to predict fluxes and photon indexes. We adopt log-uniform priors for $\rm E_{kin}$, $\rm n$, $\rm \epsilon_{e}$ and $\rm \epsilon_{B}$ and uniform priors for  $\rm p$, $\rm \theta_{j}$ and $\rm A_{R}$ (see Table \ref{tab:mcmc_values}).\\
We sample the posterior probability density through MCMC using the \textsc{emcee} python package \citep{Foreman-Mackey2013}, employing $N_{walk}=12$ walkers for $N_{iter}=500000$ iterations. The prior used for the MCMC and the results of the fit are reported in Table \ref{tab:mcmc_values}, while the corner plot with marginalized posterior distributions for each parameter is shown in Fig.~\ref{fig:mcmc_corner}.\\

\bibliographystyle{aasjournal} 
\bibliography{references} 

\end{document}